\documentclass[twocolumn,showpacs,preprintnumbers,amsmath,amssymb]{revtex4}

\usepackage{graphicx}
\usepackage{dcolumn}
\usepackage{bm}

\begin{document}

\preprint{APS/123-QED}

\title{Degenerate states of  narrow semiconductor  rings in the  presence
of spin orbit coupling: Role of time-reversal and large   gauge transformations }
\author{S.-R. Eric Yang\footnote{ eyang@venus.korea.ac.kr}}
\affiliation{Physics Department, Korea  University, Seoul Korea }


\begin{abstract}
The electron Hamiltonian of narrow semiconductor  rings with the Rashba and Dresselhaus spin orbit terms  
is invariant under time-reversal operation followed by a large gauge transformation.   
We find that all the eigenstates are doubly degenerate when integer or half-integer
quantum fluxes thread the quantum ring.  
The wavefunctions of  a degenerate pair are related to each other by the symmetry 
operation.  These results are valid even in the presence of a disorder potential.
When the Zeeman term is present only some of these degenerate levels anticross.
\end{abstract}

\maketitle

\section{Introduction}

\begin{figure}[hbt]
\begin{center}
\includegraphics[width = 0.42 \textwidth]{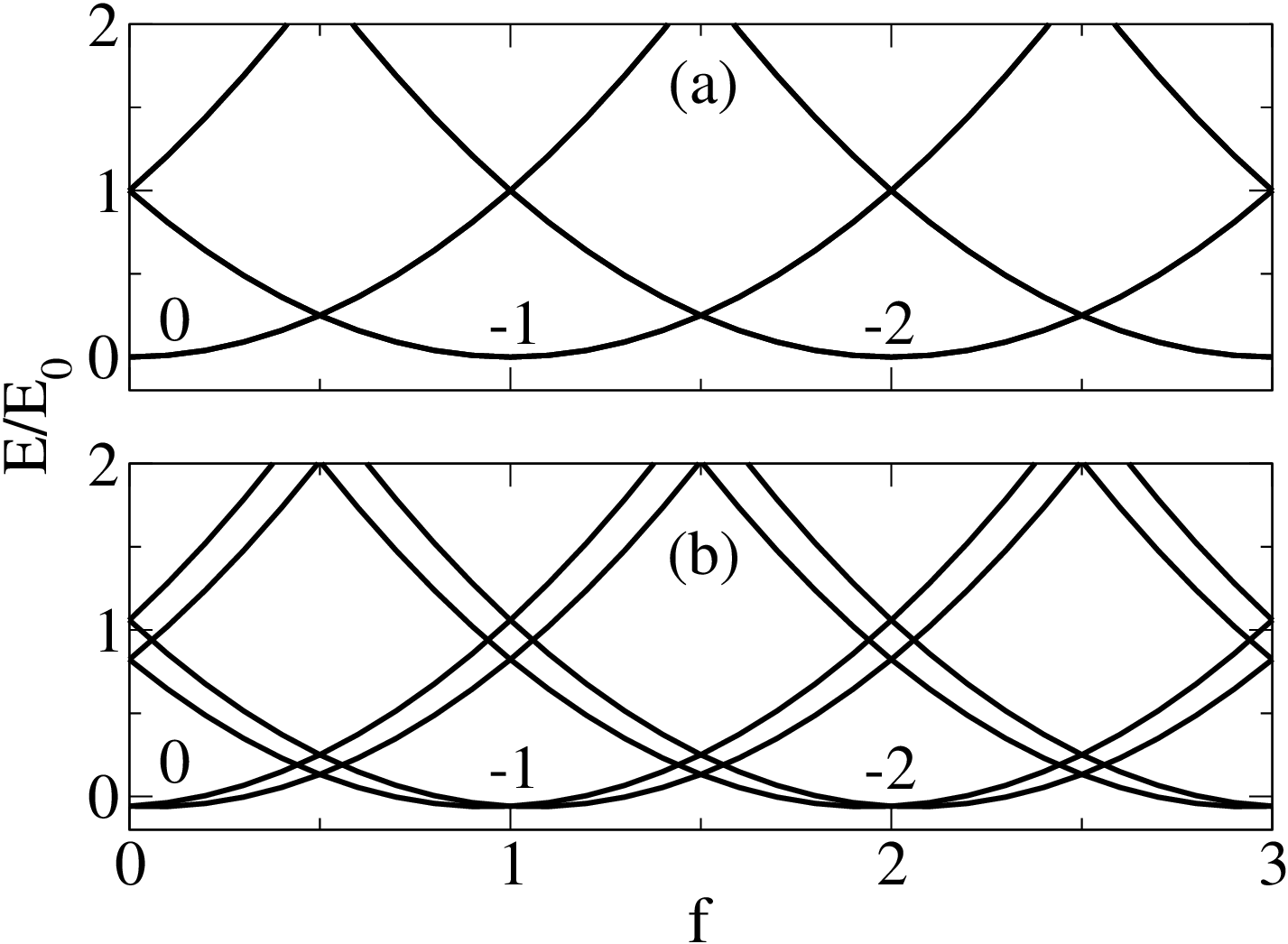}
\caption{(a)
The single electron energy of an ideal one-dimensional ring threaded by a magnetic flux $\Phi$ is 
$E_{n}(f)=E_0(n+f)^2$,  
where  $E_0$ is the  energy scale $E_0=\frac{\hbar^2}{2m^*R^2}$ 
with the electron effective mass  $m^*$ and the radius of the ring  $R$.
The electron with this energy has the wavefunction   $\Psi(\phi)=\frac{1}{\sqrt{2\pi}}e^{in\phi}$, where $n$ is the z-component of
the angular momentum.
Each  energy curve is spin degenerate and parabolic.   It takes  zero value  at integer values $f=-n$.
The numbers $0,-1,-2$ near the curves
give the   z-component of angular momentum of the electron states.
(b)  Energy levels of a narrow  ring of (a) in the presence of the Rashba term  which lifts 
spin degeneracy. The energy scale associated with the Rashba constant is set to  $E_R=0.5E_0$. 
The numbers near the curves
label the energy curves; they are no longer  z-components of angular momentum since the rotational symmetry is broken.
At integer and half integer values of $f$ the energy levels are doubly degenerate.
}
\label{fig:spectrum}
\end{center}
\end{figure}

Energy spectrum of  a quantum   ring threaded by  a magnetic flux 
is important since several interesting  properties of the ring depend on it\cite{Wash, Gefen}.
The energy spectrum is periodic with the period equal to the 
quantum unit of flux  $\Phi_0=hc/e$, where  the elementary charge  $e>0$.
This effect is  a consequence of the Aharonov Bohm effect.
The  angular momentum of an electron  in the groundstate of  a quantum ring 
in  a magnetic field can be non zero, as can be seen in the energy spectrum of Fig.\ref{fig:spectrum}(a).
This fact is closely related to the fascinating
physics of the  persistent currents in  mesoscopic rings\cite{Buttiker,Wash, Gefen}.  
The change of  groundstate angular momentum about the quantization axis from $0$ to $-1$ is a decisive
feature distinguishing quantum rings from quantum dots.
Such angular momentum transitions were  observed in self assembled quantum rings\cite{Lorke}.
Effects of spin-orbit terms  have been studied  in mesoscopic rings\cite{Meir,Morp,Aronov}.
Currently there is renewed interest in them in semiconductor quantum rings.
This is because  electron spin may be controlled by  spin-orbit terms.  Such a control  would be valuable
for spintronics, quantum information, and spin qubits\cite{Aws}.
Recently several  effects of spin orbit coupling on optical\cite{Warburton,Bayer} and transport\cite{Souma,Frust, Mol,Cap}
properties of
semiconductor quantum  rings have been investigated. 
A spin filter\cite{Nitta} and  a qubit\cite{Foldi} have been also proposed.
In this paper we investigate the energy degeneracy  of  self-assembled semiconductor rings
on the magnetic flux in the presence of spin orbit interactions. 
We show how {\it large gauge}  and {\it time reversal} transformations can be used to
determine the  degenerate properties of
the single electron
energy spectrum of such a system.

Before we give a summary of our main results we comment on some basic properties
of large gauge transformations, spin orbit terms in semiconductors, and time reserval operation.
Consider an  ideal one-dimensional 
ring without spin orbit coupling and with the radius much larger than the width of the ring.
Suppose it is  threaded by a magnetic flux $\Phi$ (the direction of the magnetic field is chosen to be along the z-axis).
It  has  rotational invariance about the z-axis that goes through the center of the ring.
In addition it is invariant under a {\it large} gauge transformation, which transforms wavefunctions as follows. 
When the flux is increased from $\Phi$  to $\Phi+\Delta\Phi$ the electron wavefunction changes as
\begin{eqnarray}
\Psi'=e^{-i\Delta f\phi}\Psi,
\end{eqnarray}
where 
{\it only} integer values of $\Delta f=\Delta\Phi/\Phi_0$ are allowed\cite{Yoshioka}.  
The azimuthal angle of the cylindrical 
coordinate system is $\phi$.
Although  time reversal symmetry is broken when  a flux  threads the ring
spin invariance is present.
Each energy level is thus doubly spin degenerate.
However, when the dimensionless flux 
$f=\Phi/\Phi_0$ is a half-integer or integer each energy level can be  4 times degenerate, as shown in  Fig.\ref{fig:spectrum}(a).

In semiconductors when the external 
confinement potential breaks the inversion symmetry the Rashba spin orbit term is relevant.
When  the crystal potential itself breaks the inversion symmetry the Dresselhaus  spin orbit term is relevant.
In II-VI semiconductors
the Rashba term is expected to be larger than the
the Dresselhaus coupling. In III-V semiconductors, such as GaAs, the
opposite is true\cite{Rashba}. 
Spin orbit terms break both spin and angular momentum symmetries
and the double degeneracy of the ideal one-dimensional ring is  broken.

In the absence of a vector potential and the Zeeman term   the spin-orbit  Hamiltonian is invariant under 
time reversal symmetry:
$\vec{k}\rightarrow -\vec{k}$ and $\vec{S}\rightarrow -\vec{S}$, where $\vec{k}$ and $\vec{S}$ are the wavevector
and spin of an electron.
The time reversal operator is 
$K=-i \sigma_y C$,
where the operator $C$ stands for complex conjugation and $\sigma_y$ is one of the  Pauli spin matrices  $\sigma_{x,y,z}$. 
If the Hamiltonian is invariant under time reversal  operation  $  \Psi \rightarrow   \Psi' $ then 
the wavefunctions   $\Psi$ and $\Psi'$ 
are degenerate and   orthonormal.
Note that $K^2 \Psi =- \Psi $.
This symmetry is the origin of Kramers' double degeneracy in quantum dots\cite{Yang}

We explore the interplay between large gauge transforation, time reversal operation, and
spin orbit coupling in semiconductor rings.
We show in the presence of the Rashba and/or Dresselhaus terms  
that, although spin double degeneracy is lifted,  all the energy levels at integer and half integer values of $f$
are {\it doubly} degenerate, see Fig.\ref{fig:spectrum}(b).
This is because at these   particular values of  $f$  a large gauge transformation  followed by  time reversal operation\cite{Yang}
is a {\it symmetry operation} of the Hamiltonian.  This result is valid even in the presence of a disorder potential and
in non-circular  rings.
The wavefunctions of  a degenerate pair are related to each other by the symmetry 
operation.
When only the Rashba term  is present analytical solutions of  eigenstates are possible. 
In this case we find, as  shown in  Fig.\ref{fig:spectrum}(b), that  the energy curve 
$E_n(f)$ of the ideal ring splits  into  curves $E_{m,L}(f)$
and $E_{n,U}(f)$, where $m$ is an integer.
As a function of $f$ the energy $E_{m,L}(f)$ or $E_{n,U}(f)$ describes a parabola.
We find that degeneracy occurs when
$E_{m,L}(f)=E_{n,U}(f)$, which leads to
\begin{eqnarray}
n=-m-2f .
\end{eqnarray}
Since $m$ and $n$ are both integers $f$ must be half integers or integers.
Unlike the previous case of the ideal ring the total degeneracy is now only two.
This relation between $m$ and $n$ can be understood in terms of time reversal and gauge transformations, as we demonstrate
below.
When the Zeeman term is present some  degenerate level anticross,  as shown in  Fig.\ref{fig:Zeeman}.

\section{Ideal one-dimensional ring}

Before we analyse the narrow semiconductor rings it would be instructive to review the basic properties of the
ideal one-dimensional ring\cite{Buttiker} in the absence of spin orbit coupling. 
The periodic nature  of the energy spectrum of an ideal one-dimensional ring 
as a function of $f$ can be understood by considering invariance under large gauge transformations.
Consider an electron moving in an ideal one-dimensional ring with radius $R$ on  the x-y plane.   
The Hamiltonian is  $H_0=\frac{1}{2m^*}(\vec{p}+\frac{e}{c} \vec{A})^2$, where $m^*$ and $\vec{A}$ are
the effective mass of the electron and the vector potential.
The electron is under the influence of the vector potential due to a  solenoid at the
center of the ring.
The vector potential given by $\vec{A}=A_{\phi}\hat{\phi}$, where
$A_{\phi}=\frac{\Phi}{2\pi R}$ and  $\Phi$ is the total magnetic flux through the ring.
Without spin orbit terms the Hamiltonian commutes with spin operator and  each energy level is {\it doubly spin degenerate}.
The wavefunction $\Psi(\phi)$ satisfies
\begin{eqnarray}
\frac{\hbar^2}{2m^*R^2}(\frac{1}{i}\frac{\partial}{\partial \phi}+ f)^2\Psi=E\Psi,
\end{eqnarray}
The solutions $\Psi(\phi)$  are  given by $\frac{1}{\sqrt{2\pi}}e^{in\varphi}$, where $n$ is an integer.
A gauge transformation
$\vec{A}'=\vec{A}+\nabla \chi(\vec{r})$ leads to the transformed wavefunction
   $\Psi'=e^{-\frac{e}{\hbar c}\chi(\vec{r})}\Psi$.
When an integral multiple of quantum unit of flux is added to the ring 
the wavefunction satisfies
\begin{eqnarray}
\frac{\hbar^2}{2m^*R^2}(\frac{1}{i}\frac{\partial}{\partial \phi}+f+\Delta f)^2\Psi'=E\Psi'.
\end{eqnarray}
The wavefunctions after and before this process are related by 
\begin{eqnarray}
\Psi'=e^{-i\Delta f\phi}\Psi.
\end{eqnarray}
Boundary condition
$\Psi'(\phi)=\Psi'(\phi+2\pi n)$  implies
that $\Psi(0)=e^{-i\Delta f 2\pi} \Psi(2\pi)$.
Since the wavefunction must be single valued 
{\it not}  all gauge transformations are allowed.  Only
those with integer $ \Delta f$ are possible.
It should be noted that  gauge transformations 
that change the flux  {\it continuously}  from zero to $\Phi_0$ do {\it not} exist.
The only possible  gauge transformations are  those that add an integral multiple of  $\Phi_0$\cite{Yoshioka}, i.e.,  
only   {\it large} gauge transformations are possible.
Under a large gauge transformation the wavefunction transforms 
\begin{eqnarray}
\Psi'=e^{-i\Delta f\phi}\Psi=e^{i(n-\Delta f)\phi}
\end{eqnarray}
The angular momentum has decreased by $\Delta f$. 
The gauge invariance requires that the energies before and after are equal:   $E_{n}(f)=E_{n-\Delta f}(f+\Delta f)$.
This should be contrasted with the  adiabatic addition of flux from zero to one unit of quantum flux,
which changes the electron 
energy from $E_0(n+f)^2$ to $E_0(n+f+1)^2$ while the wavefunction
remains unchanged.

\section{Model Hamiltonian, symmetry operations and degenerate solutions}

Let us now consider semiconductor rings with spin orbit terms.
The electron is under the influence of the vector potential due to a  solenoid at the
center of the ring.
The total Hamiltonian is $H=H_0+H_R+H_D+H_Z$.
The first part is $H_0=\frac{\hbar^2\vec{\Pi}^2}{2m^*}+U(r)+V(z)$, where $U(r)$ is the radial potential energy
and $V(z)$ is the Rashba confinement potential.
If the radius of the solenoid is equal or larger than the radius of the ring 
the electron experiences a magnetic field and the Zeeman term should be included:
$H_Z=\frac{1}{2}g_0\mu_B\sigma_z B$, where $g_0$, $\mu_B$,  and $B$ are, respectively, the effective g-factor,
the Bohr magneton, and the magnetic field.
$\Pi_{x,y}$ are the kinematic momentum operators with $\Pi_{x,y}=k_{x,y}+\frac{e}{\hbar c}A_{x,y}$
and  $k_x=\frac{1}{i}\frac{d}{dx}$.  
The x and y components of the vector potential are $A_x=-A_{\phi}sin\phi$  and $A_y=A_{\phi}cos\phi$.
The Rashba spin orbit term is 
\begin{eqnarray}
H_\mathrm{R}=c_\mathrm{R} \left( \sigma_x \Pi_y -\sigma_y \Pi_x \right).
\end{eqnarray}
The constant $c_R$ {\it depends on the external electric field} $E$ applied along the z-axis.
The Dresselhaus spin orbit term is
\begin{eqnarray}
H_\mathrm{D}=c_\mathrm{D}\left( 
 \sigma_x \Pi_x \left(\Pi_y^2-\Pi_z^2 \right) 
+ \sigma_y \Pi_y \left(\Pi_z^2-\Pi_x^2  
\right)\right).
\label{Dresselhaus}
\end{eqnarray}
There is another term of the 
form  $ \sigma_z \langle \Pi_z\rangle \left(\Pi_x^2-\Pi_y^2 \right) $ in the  Dresselhaus spin orbit term
but it vanishes  since the expectation value of the first subband wavefunction $f(z)$  along  z-axis
$\langle \Pi_z\rangle =\langle f(z)|k_z|f(z)\rangle =0$.
The  constant $c_D$  represents breaking of inversion symmetry by the crystal  in zinc blende structures. 
The  confinement  potentials along the z-axis and radial direction are assumed to be sufficiently strong 
that
only the lowest energy subbands are relevant. 
The total electron wavefunction in a narrow semiconductor ring can be written as 
$\Phi(\vec{r})=f(z)R(r)\Psi(\phi)v(\vec{r})$.  The lowest subband wavefunction along the radial direction is $R(r)$.
The conduction band Bloch wavefunction is $v(\vec{r})$. In the following
when the electron wavefunction is written only the azimuthal part $\Psi(\phi)$ will be  shown and 
other  wavefunctions will be suppressed.

The azimuthal part of the wavefunction can be written as
\begin{eqnarray}
\Psi=\sum_{m=-\infty}^{\infty}c_{m\uparrow}e^{im\phi}|\uparrow>+\sum_{n=-\infty}^{\infty}c_{n\downarrow}e^{in\phi}|\downarrow>,
\end{eqnarray}
or in spinor notation
$\Psi=
\left(
\begin{array}{c}
F_{\uparrow}(\phi) \\
F_{\downarrow}(\phi)
\end{array}
\right)$,
where $F_{\uparrow}(\phi)=\sum_{m=-\infty}^{\infty}c_{m\uparrow}e^{im\phi}$   
and $F_{\downarrow}(\phi)=\sum_{n=-\infty}^{\infty}c_{n\downarrow}e^{in\phi}$.
Under  time reversal operation\cite{Yang}  $  \Psi \rightarrow   \Psi' $, i.e.,
\begin{eqnarray}
\left(
\begin{array}{c}
F_{\uparrow}(\phi) \\
F_{\downarrow}(\phi)
\end{array}
\right)
\rightarrow 
\left(
\begin{array}{c}
-F_{\downarrow}^*(\phi) \\
F_{\uparrow}^*(\phi)
\end{array}
\right),
\end{eqnarray}
or
\begin{eqnarray}
\Psi'=K  \psi
=-\sum_{m }e^{-im\phi}c_{m \downarrow}^* |  \uparrow \rangle
+\sum_{n } e^{-in\phi}c_{n \uparrow}^* |  \downarrow \rangle.
\end{eqnarray}
We have suppressed the Bloch wavefunction of the conduction band in applying the time reversal
operator since it is unaffected by the operator $K$. Our wavefunctions are 
all effective mass wavefunctions  and only the  conduction band Bloch
wavefunction at $\vec{k}=0$ is relevant.
Note that time reversal symmetry is {\it absent} when a vector potential is present.

Suppose that a solenoid with  an integer or half integer  fluxes, $k\Phi_0$,  threads the quantum ring. 
The vector potential of the solenoid is $\vec{A}$. Consider a large gauge transformation that flips the direction of 
the vector potential: $\vec{A}\rightarrow \vec{A}-2\vec{A}=-\vec{A}$. This operation is equivalent to the subtraction
of the flux  $2k\Phi_0$.
If the flux $2k\Phi_0$ is {\it subtracted} first and time reversal operation is applied next we find 
that the wavefunction transforms as follows:
\begin{eqnarray}
\left(
\begin{array}{c}
F_{\uparrow}(\phi) \\
F_{\downarrow}(\phi)
\end{array}
\right)
\rightarrow 
e^{i2k\phi}\left(
\begin{array}{c}
F_{\uparrow}(\phi) \\
F_{\downarrow}(\phi)
\end{array}
\right)
\rightarrow
e^{-i2k\phi}\left(
\begin{array}{c}
-F_{\downarrow}^*(\phi) \\
F_{\uparrow}^*(\phi)
\end{array}
\right).\nonumber\\
\end{eqnarray}
Note that this procedure is equivalent to
applying  time reversal operation  first and {\it adding} the flux $2k\Phi_0$ next: 
\begin{eqnarray}
\left(
\begin{array}{c}
F_{\uparrow}(\phi) \\
F_{\downarrow}(\phi)
\end{array}
\right)
\rightarrow
\left(
\begin{array}{c}
-F_{\downarrow}^*(\phi) \\
F_{\uparrow}^*(\phi)
\end{array}
\right)
\rightarrow
e^{-i2k\phi}\left(
\begin{array}{c}
-F_{\downarrow}^*(\phi) \\
F_{\uparrow}^*(\phi)
\end{array}
\right).\nonumber\\
\end{eqnarray}
From the Hamiltonian it can be seen that under the gauge transformation, 
$\vec{A}\rightarrow -\vec{A}$, and the   time reversal operation, 
$\vec{k}\rightarrow -\vec{k}$ and $\vec{S}\rightarrow -\vec{S}$,
the electron Hamiltonian is invariant in the absence of the Zeeman term. 
The angular momentum components of the wavefunctions before and after these  operations are related as follows:
\begin{eqnarray}
\Psi=\left(
\begin{array}{c}
\sum_{m} a_m e^{im\phi} \\
\sum_{n} b_n e^{in\phi}
\end{array}
\right)
\end{eqnarray}
and 
\begin{eqnarray}
\Psi'=\left(
\begin{array}{c}
\sum_{m} A_{m} e^{im\phi} \\
\sum_{n} B_{n} e^{in\phi}
\end{array}
\right),
\end{eqnarray}
where  $A_m=- b_{-m-2k}^*$   and $B_n=a_{-n-2k}^*$.  These wavefunctions are degenerate.
Unlike the  case of the ideal one-dimensional ring  the total degeneracy is now only two.
The result given in this section is valid  in the absence of the Zeeman term.
However, a disorder potential and non-circular ring will not break 
the presence of degeneracies becuase the relevant potential $V_D(\vec{r})$ of these systems 
is invariant under the gauge transformation,
$\vec{A}\rightarrow -\vec{A}$, and the   time reversal operation,
$\vec{k}\rightarrow -\vec{k}$ and $\vec{S}\rightarrow -\vec{S}$.

\section{Analytic solutions when the  Rashba spin orbit is present}
A narrow semiconductor  ring with  the  Rashba spin orbit  term can be solved exactly even in the presence of the Zeeman term.
The radial wavefunction $R(r)$ is taken to be a Gaussian and 
the correct   Hamiltonian\cite{Meijer}  is
\begin{eqnarray}
\left(
\begin{array}{cc}
E_0 (\frac{1}{i}\frac{\partial}{\partial\phi}+f)^2+z   &   E_Re^{-i\phi}(f-1/2+\frac{1}{i}\frac{\partial}{\partial\phi}) \\
E_R e^{i\phi}(f+1/2+\frac{1}{i}\frac{\partial}{\partial\phi})            & E_0 (\frac{1}{i}\frac{\partial}{\partial\phi}+f)^2-z
\end{array}
\right).\nonumber\\
\label{matrix_Hamil}
\end{eqnarray}
Here
 $E_R =\frac{c_R}{R}$.   
Note that  the Zeeman term can be written as $z=g\mu_0 B/2= z_0 f$, where   $z_0=g\mu B_0/2$ and    $B_0=\Phi_0/\pi R^2$.
The eigenstates are  of the form\cite{Aronov,Frust,Mol} 
\begin{eqnarray}
\Psi_m=
\left(
\begin{array}{c}
a_m e^{im\phi} \\
b_{m+1} e^{i(m+1)\phi}
\end{array}
\right).
\end{eqnarray}
The coefficients $a_m$ and $b_{m+1}$ satisfy the matrix equation 
\begin{eqnarray}
\left(
\begin{array}{cc}
E_0 p^2 +z  &   E_R(p+1/2) \\
E_R (p+1/2)            & E_0 (p+1)^2-z
\end{array}
\right)
\left(
\begin{array}{c}
a_m \\
b_{m+1} 
\end{array}
\right)=E\left(\begin{array}{c}
a_m \\
b_{m+1}
\end{array}
\right)
\nonumber\\
\label{matrix_Hamil}
\end{eqnarray}
with two eigenvalues   
\begin{eqnarray}
E_{\pm}=
\frac{1}{2}E_0(1+2p+2p^2)\pm\frac{1}{2}X(E_R,p,z),\nonumber\\
\end{eqnarray}
where
$p=f+m$  and
\begin{eqnarray}
&&X(E_R,p,z)\nonumber\\
&=&[(E_0^2+E_R^2)(1+2p)^2+4z^2-4E_0z(1+2p)]^{1/2}.\nonumber\\
\end{eqnarray}
The  
corresponding eigenvectors are
\begin{eqnarray}
\Psi_{m,\pm}=\left(
\begin{array}{c}
a_{m}^{\pm} e^{im\phi}\\
b_{m+1}^{\pm}e^{i(m+1)\phi}
\end{array}
\right),
\label{eigenvec}
\end{eqnarray}
where 
\begin{eqnarray}
\left(
\begin{array}{c}
a_{m}^{\pm}\\
b_{m+1}^{\pm}
\end{array}
\right)
=
\frac{1}{N_{\pm} }
\left(
\begin{array}{c}
-E_0-2E_0p+2z\pm X \\
E_R|1+2p|
\end{array}
\right)
\end{eqnarray}
with the normalization factors 
\begin{eqnarray}
N_{\pm}=[E_R^2(1+2p)^2+(E_0(1+2p)-2z\mp X)^2]^{1/2}.\nonumber\\
\end{eqnarray}
\
\
\begin{figure}[hbt]
\begin{center}
\includegraphics[width = 0.4 \textwidth]{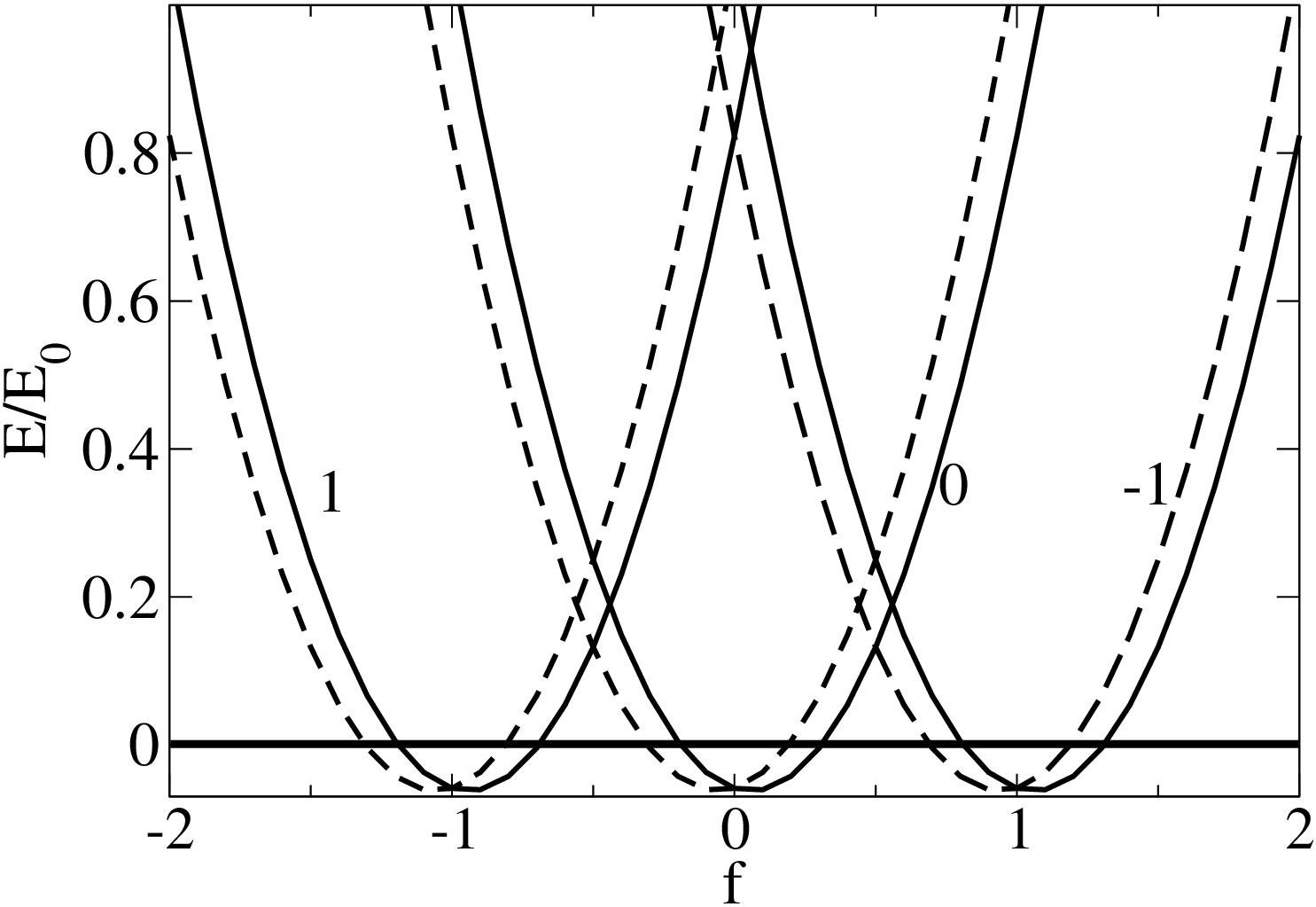}
\caption{Energy levels of a  one-dimensional ring in the presence of the Rashba term
$E_R=0.5E_0$.
Solid lines  from right to left: $E_{-1,L}(f)$,  $E_{0,L}(f)$, and  $E_{1,L}(f)$.
Dashed lines from right to left: $E_{-1,U}(f)$,  $E_{0,U}(f)$, and  $E_{1,U}(f)$.}
\label{fig:setup}
\end{center}
\end{figure}

\subsection{Absence of the Zeeman term}
When the Zeeman term is zero the expression for the eigenvalues simplifies 
\begin{eqnarray}
E_{\pm}&=&\frac{1}{2}E_0(1+2p+2p^2)\nonumber\\
&\pm&\frac{1}{2}\sqrt{(E_0^2+E_R^2)(1+2p)^2}.\nonumber\\
\end{eqnarray}
These energies are not smooth functions of $f$.
They  can be combined into smooth functions $E_{m,L}(f)$ and $E_{m,U}(f)$ that are given by
\begin{eqnarray}
E_{m,L}(f)&=&\frac{1}{2}E_0(1+2p+2p^2)
-\frac{(1+2p)}{2}\sqrt{(E_0^2+E_R^2)}\nonumber\\
&=&E_0(m+f-F_{-})(m+f-F_{+})
\end{eqnarray}
and
\begin{eqnarray}
E_{n,U}(f)&=&\frac{1}{2}E_0(1+2k+2k^2)
+\frac{(1+2k)}{2}\sqrt{(E_0^2+E_R^2)}\nonumber\\
&=&E_0(n+f-G_{-})(n+f-G_{+}),
\end{eqnarray}
where $k=n-1+f$.
Note 
\begin{eqnarray}
F_{\pm}=\frac{1}{2}(\sqrt{1+(E_R/E_0)^2}-1)\pm\frac{1}{2}E_R/E_0\nonumber\\
G_{\pm}=-\frac{1}{2}(\sqrt{1+(E_R/E_0)^2}-1)\pm\frac{1}{2}E_R/E_0.
\end{eqnarray}
The corresponding eigenvectors are
\begin{eqnarray}
\Psi_{m,L}=\left\{
\begin{array}{cc}
\Psi_{m,+}& \textrm{for} \ 1+2(m+f)<0\\
\Psi_{m,-}&\textrm{for} \ 1+2(m+f)>0
\end{array}
\right.
\end{eqnarray}
and
\begin{eqnarray}
\Psi_{n,U}=\left\{
\begin{array}{cc}
\Psi_{n-1,-}&\textrm{for} \ 1+2(n-1+f)<0\\
\Psi_{n-1,+}&\textrm{for} \  1+2(n-1+f)>0.
\end{array}
\right.
\end{eqnarray}
When $E_R=0$ the constants $F_{\pm}=G_{\pm}=0$ and  $E_{m,L}(f)=E_{m,U}(f)=E_0(m+f)^2$.   
As a function of $f$ the energies  $E_{m,L}(f)$ and  $E_{n,U}(f)$ describe parabola, as shown in  Fig.(\ref{fig:setup}).
The roots of $E_{m,L}(f)$ are $f_{-}=F_{-}-m$ and $f_{+}=F_{+}-m$.
The difference between them is $E_R/E_0$.  The roots of $E_{n,U}(f)$ are $f_{-}=G_{-}-n$ 
and $f_{+}=G_{+}-n$.
Again the difference between them is $E_R/E_0$.

Degeneracy at   {\it half integer or integer} values of $f$ occurs when the condition
 $E_{m,L}(f)=E_{n,U}(f)$ is satisfied.  This  leads to 
\begin{eqnarray}
n=-m-2f .
\end{eqnarray}
At $f=1/2$, for example, the  pair represented by the quantum numbers 
$(m,n)=(0,-1)$
is degenerate:  $E_{-1,U}(1/2)=E_{0,L}(1/2)$.   The corresponding 
degenerate wavefunctions are 
\begin{eqnarray}
\Psi_{-1,U}=\Psi_{-2,-}=
\left(
\begin{array}{c}
a e^{-2i\phi} \\
b e^{-i\phi}
\end{array}
\right)
\end{eqnarray}
and 
\begin{eqnarray}
\Psi_{0,L}=\Psi_{0,-}=
\left(
\begin{array}{c}
b  \\
-a e^{i\phi}
\end{array}
\right),
\end{eqnarray}
where a and b are real constants.
They are  related to each other by the  subtraction of   one   unit of quantum flux and 
the application  of time-reversal operation: 
 $\Psi_{-1,U}=K e^{i\phi}\Psi_{0,L} $.
Note that in order to obtain this result we have multiplied
 the expression of $\Psi_{0,L}$, given  by
 Eq.(\ref{eigenvec}), with a phase factor.

\subsection{Presence of  Zeeman effect}

\begin{figure}[hbt]
\begin{center}
\includegraphics[width = 0.4 \textwidth]{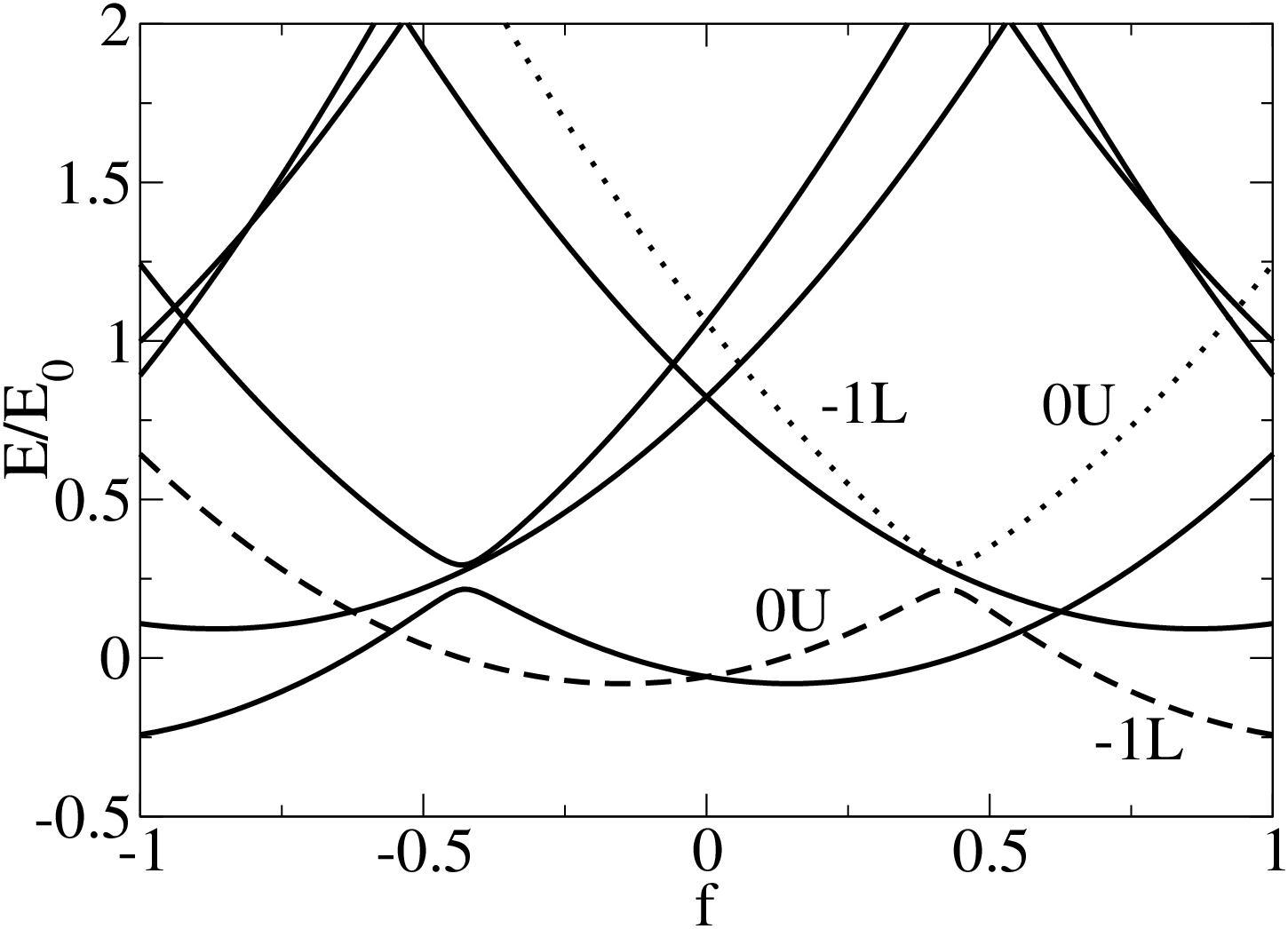}
\caption{Energy levels of a  one-dimensional narrow ring in the presence of the Rashba interaction
and the Zeeman term $E_R=0.5E_0$.
 Anticrossing of the  energy curves $E_{0,U}$   and $E_{-1,L}$ near $f=0.5$.
The labels on the dashed and dotted energy lines, $0U$ and  $-1L$,
indicate the correct quantum numbers in the absence of the Zeeman term.}
\label{fig:Zeeman}
\end{center}
\end{figure}

Suppose the electron experiences an external magnetic field along the z-axis with the  vector potential  
$\vec{A}=\frac{B}{2}(-y,x)$.
To illustrate the effect of the Zeeman term we calculate the energy spectrum for 
$In_xGa_{1-x}As$ with the physical parameters $g=-4$ and  
$m^*=0.05m$.  For this semiconductor the Rashba term is the dominant spin orbit coupling.
We assume that the radius is much larger than the ring width.
For the radius of the ring  $R=14nm$ the energy scale is  $E_0=3.9meV$
and 
$B[T]=6.70 f$.  
The Zeeman term couples the degenerate states $\Psi_{0,U}$  and   $\Psi_{-1,L}$ since 
 $<\Psi_{0,U}|H_Z|\Psi_{-1,L}>=<\Psi_{-1,-}|H_Z|\Psi_{-1,-}>$ is non zero. But 
it does not couple  $\Psi_{-1,U}$ and 
$\Psi_{0,L}$ 
since $<\Psi_{0,L}|H_Z|\Psi_{-1,U}>=<\Psi_{0,-}|H_Z|\Psi_{-2,-}>$ is zero.
This is the reason why there is only one anticrossing near, for example,  $f=1/2$, as can be 
see in Fig.(\ref{fig:Zeeman}).
The value of degenerate energy is $E_0/4$ in the absence of the Zeeman term.
Note that the degeneracy between   $\Psi_{-1,L}$ and  $\Psi_{-1,U}$  at $f=1$ is lifted 
unlike the degeneracy between   $\Psi_{0,L}$ and  $\Psi_{0,U}$  at $f=0$.

\
\
\
\

\section{Discussions}

We have argued that the electron Hamiltonian of narrow semiconductor  rings with the Rashba and Dresselhaus spin orbit terms  
is invariant under time-reversal operation followed by a large gauge transformation, provided that the Zeeman effect
is absent.   
We find that all the eigenstates are doubly degenerate when integer or half-integer
quantum fluxes thread the quantum ring.  
The wavefunctions of  a degenerate pair are related to each other by the symmetry 
operation.
These qualitative results are valid both for II-VI and III-V semiconductor rings.
A disorder potential and non-circular ring will not break the relevant  symmetry
and the presence of degeneracy.
We have obtained analytical and simple expressions for the energy spectrum 
when the Rashba term is dominant.
Some of degenerate energy levels anticross in the presence of  the Zeeman term.

Since the Zeeman effect
complicates the energy spectrum to some degree, as shown  in Fig.(\ref{fig:Zeeman}), 
semiconductor rings with a rather small  Zeeman effect, such as GaAs rings, 
are  more suitable for experimental study of the 
level degeneracy. However, in this case, a considerable numerical work is needed
to compute a quantitatively correct  energy spectrum due to the presence of the Dresselhaus term.
Measurement of conductance oscillations\cite{Aronov,Souma,Frust,Mol,Cap} or
optical emission lines\cite{Bayer} may reveal the level degeneracy.

Another suitable ring system to investigate 
is a superconductor and semiconductor hybrid system\cite{Berciu}.
Each  vortex of the  Abrikosov lattice of the superconductor 
has half of the quantum unit flux, and
a semiconductor ring under the superconductor  can trap this flux\cite{Byers}.
The distance between the vortices must be  larger than the diameter of the semiconductor ring, and
this may be achieved by controlling  the strength of the magnetic field.
The advantage of such a system is that the Zeeman effect is absent since 
only the central part of the ring is threaded  with  half integer flux.  
When an InAs semiconductor
is used, for example, the Rashba term is dominant, and  our  analytical results
should describe well the energy spectrum of such a  ring.

It is desirable to calculate numerically the energy spectrum quantitatively when   both the Rashba and Dresselhaus terms
are present in addition to the Zeeman term.  It may be worthwhile to investigate the effect of the finite width
of the ring
\cite{Bee}.  The double degeneracy at an integer or half integer flux may be used to generate non-Abelian Berry 
phases, see Ref.19.

\begin{acknowledgments}
This work was  supported by grant No. R01-2005-000-10352-0 from the Basic Research Program
of the Korea Science and Engineering
Foundation and by Quantum Functional Semiconductor Research Center (QSRC) at Dongguk University
of the Korea Science and Engineering
Foundation. 
\end{acknowledgments}


\begin{references}

\bibitem{Wash} S. Washburn and R.A. Webb, Adv. Phys.  {\bf 35}, 395 (1986);
A.G. Aronov and Y. V. Sharvin, Rev. Mod. Phys. {\bf 59}, 755 (1987).

\bibitem{Gefen} Y. Gefen, Y. Imry, and M. Y. Azbel,  Phys. Rev. Lett. {\bf 52}, 129 (1984).




\bibitem{Buttiker}M. Buttiker, Y.Imry, and R. Landauer, Phys.Lett, {\bf 96A}, 365 (1983).

\bibitem{Lorke} A. Lorke, R.J. Luyken, A. O. Govorov, J. Kotthaus, J.M. Garcia, and P.M. Petroff,
Phys. Rev. Lett. {\bf 84}, 2223 (2000).


\bibitem{Meir} Y.Meir, Y. Gefen, and O.Entin-Wohlman  Phys. Rev. Lett. {\bf 63}, 798 (1989).

\bibitem{Aronov}A.G. Aronov and Y.B. Lyanda-Geller, Phys. Rev. Lett. {\bf 70}, 343 (1993).

\bibitem{Morp}A.F. Morpurgo, J.P. Heida, T.M. Klapwijk, and B.J. van Wess,  Phys. Rev. Lett. {\bf 80}, 1050 (1998).

\bibitem{Aws}D.D. Awschalom, D. Loss, and N. Samarth, Semiconductor Spintronics and Quantum Computation (Springer, Berlin, 2002).


\bibitem{Warburton} R.J. Warburton, C. Schaflein, D. Haft, F. Bickel, A. Lorke, K. Karrai, J.M. Garcia, 
W. Schoenfeld, and P.M. Petroff, Nature  {\bf 405}, 926 (2000).

\bibitem{Bayer}M. Bayer, M. Korkusinski, P. Hawrylak, T. Gutbrod, M. Michel, and Forchel,  
Phys. Rev. Lett. {\bf 90}, 186801  (2003).

\bibitem{Souma} S. Souma and B. Nikoli\'{c},  Phys. Rev. B, {\bf 70} 195346 (2004).

\bibitem{Nitta}J. Nitta, F.E. Meijer, and H. Takayanagi, Appl. Phys. Lett.  {\bf 75}, 695 (1999).

\bibitem{Foldi}P. F\"{o}ldi, B. Moln\'{a}r, M.G. Benedict, and F.M. Peeters, Phys. Rev. B, {\bf 71}, 033309 (2005).



\bibitem{Frust}D. Frustaglia and K. Richter,  Phys. Rev. B, {\bf 69} 235310 (2004).

\bibitem{Mol} B. Moln\'{a}r, F.M. Peeters, and P. Vasilopouos,  Phys. Rev. B, {\bf 69} 155335 (2004).


\bibitem{Cap} R. Capozza, D. Gilliano, P. Lucignano, and A. Tagliacozzo, Phys. Rev. Lett. {\bf 95}, 226803 (2005).




\bibitem{Yoshioka}D. Yoshioka, The Quantum Hall Effect (Springer, Berlin, 1998).

\bibitem{Rashba}E. Rashba cond-mat/0507007

\bibitem{Yang} S.-R. Eric Yang and N.Y. Hwang, Phys. Rev. B, {\bf 73}, 125330 (2006).


\bibitem{Meijer}F.E. Meijer, A.F. Morpurgo, and T.M. Klapwijk, Phys. Rev. B, {\bf 66}, 033107 (2002).


\bibitem{Berciu}M. Berciu, T.G. Rappoport, and B. Janko, Nature {\bf 435}, 71 (2005).

\bibitem{Byers}N. Byers and C.N. Yang, Phys. Rev. Lett. {\bf 7}, 46 (1961).

\bibitem{Bee} C.W. Beenakker, H.van Houten, and A.A.M. Staring, Phys. Rev. B, {\bf 44}, 1657 (1991).



\end{references}
\end{document}